# Inverted rear-heterojunction GaInP solar cells using Te memory effect

Manuel Hinojosa, Iván García[*], Ignacio Rey-Stolle and Carlos Algora

*Instituto de Energía Solar - E.T.S.I. Telecomunicación, Universidad Politécnica de Madrid, Madrid, Spain*

[*]Corresponding author: igarcia@ies.upm.es

## ABSTRACT

Tellurium allows attaining heavy n-type doping levels in GaAs, which is suited to achieve very low contact resistivities in solar cells. Besides, it modifies the energy bandgap of MOVPE-grown GaInP by reducing the group-III sublattice ordering and presents a strong memory effect which induces residual n-type doping in subsequent layers, potentially detrimental to the performance of the solar cell. In this work, we present an inverted rear-heterojunction GaInP solar cell that employs a thick Te-doped GaInP layer as absorber, with a doping profile obtained exclusively by controlling the memory effect of Te coming from the preceding growth of a heavily doped GaAs contact layer. In this way, GaInP is partially disordered with the use of no additional surfactant, leading to an increase in the solar cell bandgap of around 35 meV as compared to traditional samples doped with silicon. In the proof-of-concept experimental devices developed so far, the use of a rear-heterojunction configuration and the bandgap increase results in a global open-circuit voltage enhancement of 109 mV. The photocurrent decreases by 1.32 mA/cm$^2$, mostly due to the bandgap blue-shift, with about 0.35 mA/cm$^2$ attributable to lower carrier collection efficiencies. These preliminary results are discussed by analyzing the I-V curve parameters and quantum efficiencies of a Te-doped rear-heterojunction, a Si-doped rear-heterojunction and a Si-doped front-junction solar cell. An additional advantage is that the emitter sheet resistance is reduced from 551 to 147 Ω/□, which offers potential for higher efficiencies through lower front grid shadowing factors, as demonstrated with the concentrator measurements presented.

## 1. Introduction

GaInP is a III-V semiconductor commonly used as the absorber material in the top subcell of most of high-efficiency multijunction solar cell (MJSC) architectures [1] [2] [3] [4]. The potential of this material has to do with (i) a suitable bandgap ($E_g$) for the top cell in a 3-or-4 junction solar cell when lattice-matched to GaAs or Ge and with (ii) a high electronic quality which allows to efficiently collect the minority carriers generated in the bulk by the absorption of photons. The design of the GaInP top subcell layers and its epitaxial growth process aims to achieve a subcell performance as high as possible –high collection efficiencies and low recombination currents– while making it compatible with a high overall performance of the MJSC device. For instance: the use of surfactant elements to blue-shift its bandgap is normally desired to optimize the MJSC performance but can affect the performance of the top cell [5]; the reduction of the MJSC overall series resistance involves the GaInP top cell emitter sheet resistance ($R_{she}$) design [6] or the use of very high doping levels in the front contact layer –only separated by a few nm from the top subcell active layers– to obtain low specific metal/semiconductor contact resistances ($\rho_c$) can interact with the GaInP subcell through diffusion [7], memory effect [8] or the injection of point defects [9].

When GaInP is grown in a metal-organic-vapour-phase-epitaxy (MOVPE) reactor, it typically presents some spontaneous ordering in the group-III sublattice, which lowers its $E_g$ and, consequently, reduces the open-circuit-voltage ($V_{oc}$), the transmitted light to subcells underneath and, in the end, the maximum theoretical efficiency attainable by the MJSC [10]. One common strategy to minimize such effect is based on the use of surfactant elements, such as Sb [11] [5], during the growth of the GaInP absorber, in order to promote disordering and get its bandgap as close to the ideal as possible. However, the use of Sb during the growth of a GaInP solar cell must be carefully evaluated as it has proven to reduce the lifetime of carriers [12] [13].

On another note, the lateral spreading of the current from the semiconductor towards the front metal contact occurs in the emitter and window layers of the topmost subcell. Therefore, the reduction of the top subcell $R_{she}$ is of particular interest in concentrator applications as it allows to obtain higher peak efficiencies in the trade-off between series resistance and shadowing factor that decides the front grid finger density. Although a common way to address this issue is based on highly-doped thin-emitters [6], recent works have demonstrated that rear-heterojunction designs (RHJ), with thick n-type emitters, can achieve enough minority carrier diffusion lengths to collect carriers generated far away from the junction, improve the external radiative efficiency ($\eta_{ext}$) over traditional front-junction (FJ) structures and even decrease the $R_{she}$, since the doping level is reduced by a lower factor than the emitter thickness increase, as shown by Geisz et al. with GaInP single-junction solar cell efficiencies exceeding 20% at 1 sun [14]. Another crucial parameter to take into account is the front metal-semiconductor specific contact resistance, which, in brief, determines the total metal area (i.e., the shadowing factor) that is necessary to ensure a lossless current transfer from the semiconductor to the metal. For the typical metal systems used, a high doping level in the semiconductor contact layer is required.

In an epitaxial growth process, the use of Te presents some benefits as compared to other typical n-type elements like silicon (Si), such as a non-amphoteric nature or the possibility to reach very high doping levels [15], but also involves some growth effects that have to be considered, namely: (i) Te tends to stay on the growth surface and acts as surfactant, reducing the CuPt sublattice ordering [16] or (ii) it presents a remarkable memory effect [17], which is the tendency of an element to continue incorporating into the epilayers after it has been shut-off, complicating the achievement of abrupt doping profiles.

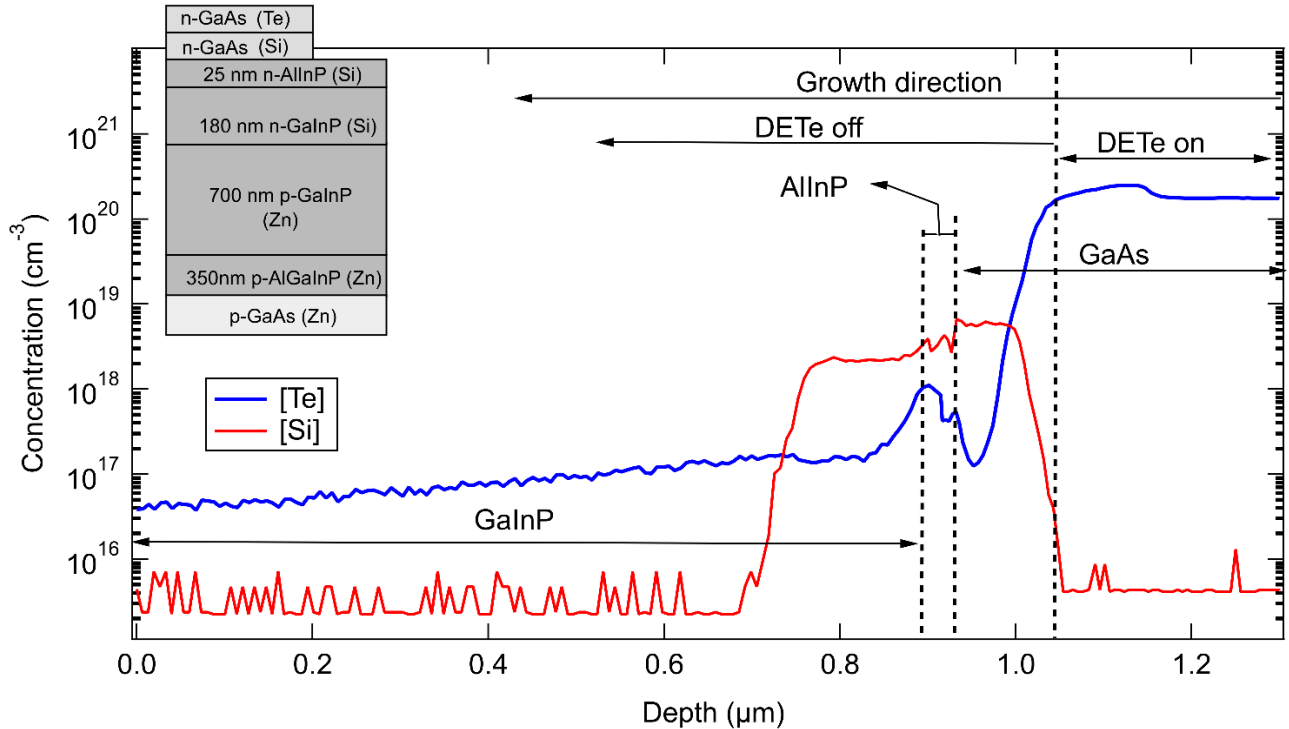

**Figure 1.** SIMS measurement of n-type dopants in a traditional inverted GaInP solar cell. Due to the memory effect, Te continues incorporating after the DETe flow is shut-off. Other elements not shown have been used to delineate the layers.

To illustrate the relevance of Te-memory effect in a MOVPE epitaxial process, Fig. 1 presents a SIMS measurement of the concentration of the n-type dopant elements used in our traditional FJ inverted GaInP solar cell. This solar cell design, sketched in the inset of Fig. 1 and detailed in next section, employs a thin Si-doped GaInP layer as emitter and a heavily Te-doped GaAs layer as contact layer, which is grown at the beginning of the structure to ensure a low specific metal contact resistance. In order to minimize the residual Te incorporation in the cell absorber, an additional Si-doped GaAs layer is placed between the AlInP window layer and the GaAs contact layer. As expected, a gradual reduction in [Te] takes place in the mentioned GaAs layer after DETe is shut off. However, [Te] increases as soon as the phosphide layers start, decreasing steadily as the growth proceeds. This appears to indicate that Te atoms staying on the growth surface are incorporated more efficiently in (Al)(Ga)InP than in GaAs leading to a residual concentration within a range of $4 \cdot 10^{16}$ to $1 \cdot 10^{17}$ atoms/cm$^3$ in the phosphide absorber layers, including the p-type base layer. Therefore, a higher Zn flow is required to achieve the nominal p-type doping level with this background n-type doping giving rise to a high compensation factor in the final Zn-doped (~$1 \cdot 10^{17}$ cm$^{-3}$) base layer.

In this work we consider tellurium (Te) both as n-type doping element and disordering surfactant in the thick emitter of a RHJ GaInP solar cell (Te-RHJ). The genesis of this study is in an effort to take advantage of the high doping levels and low diffusion coefficient of Te in GaAs to achieve a very good contact layer [18] while dealing with the memory effect explained above and illustrated in Fig. 1. Note that in these inverted designs the contact layer suffers a heavy thermal load during the growth of the rest of the multijunction structure. Using other dopants such as Se, which also allows to obtain very high doping levels, has been found to produce diffusion-related problems in these inverted structures [7]. We describe a method to grow an inverted RHJ solar cell formed by a thick n-GaInP absorber with doping obtained exclusively by modulating the memory effect of Te. This solution allows to use to our advantage the unavoidable memory effect of Te coming from a highly doped front contact layer by (i) increasing the bandgap and the $V_{oc}$ by the reduction of the spontaneous partial CuPt ordering without the use of any additional surfactant; (ii) reducing the emitter sheet resistance by the increase of the emitter thickness and average doping level and (iii) further increasing the $V_{oc}$ by using a RHJ architecture which has already demonstrated to enhance the radiative efficiency of a GaInP solar cell. The contributions to the experimental $V_{oc}$ gain corresponding to the bandgap increase and the solar cell configuration are assessed through the comparison of Te-doped RHJ samples with a Si-doped FJ and a Si-doped RHJ benchmark samples. In this work we present and analyze the first experimental results corresponding to prototype, proof-of-concept implementations of this solar cell design. Further optimizations of these devices are possible by tackling the effect of the slightly lower collection efficiency observed in the RHJ configuration, which is discussed in detail.

## 2. Experimental

A set of solar cell samples were grown on GaAs substrates with a miscut of 2º towards the (111)B plane in a horizontal low-pressure MOVPE reactor (AIX 200/4). The precursors used were $AsH_3$ and $PH_3$ for group-V, TMGa and TMIn for group-III and DETe, DTBSi and DMZn for doping elements. The semiconductor structure of the inverted Te-RHJ solar cells, sketched in Fig. 2, is based on the RHJ design mentioned above [14]. It begins with a n++ GaAs front contact layer doped using a fixed DETe concentration and growth conditions for all the samples, at values determined to obtain Te doping levels around $1 \cdot 10^{20}$ cm$^{-3}$ that allow very good metal/semiconductor specific contact resistances [18]. Then, at a fixed temperature of 675ºC, 25 nm of $Al_{0.53}In_{0.47}P$ (AlInP) and 25 nm of $Ga_{0.51}In_{0.49}P$ (GaInP) are grown while injecting DETe concentrations which vary from run to run, referred to as *phosphide Te preload flows* from now on. We test 4 different Te preloads: one where Te is flown only during the growth of the contact layer, without phosphide Te preload (Te-RHJ#1); and other three where Te is additionally injected during the first nanometers of the phosphide layers, with varying phosphide Te preloads flows (Te-RHJ#2 to #4). After the preload step, the DETe run valve is switched-off and the remainder of the GaInP absorber layer is grown being doped exclusively by the memory effect of Te. GaInP is grown using a V/III ratio of 120, a growth rate of 4.4 µm/h and molar flows of $f_{TMIn} = 9.5 \cdot 10^{-5}$ mol/min and $f_{TMGa} = 4 \cdot 10^{-5}$ mol/min. Finally, a p+ $Al_{0.25}Ga_{0.25}In_{0.49}P$ layer (AlGaInP), doped to around $1 \cdot 10^{18}$ cm$^{-3}$, forms the rear p-n junction, and a p++-GaAs layer, the back contact.

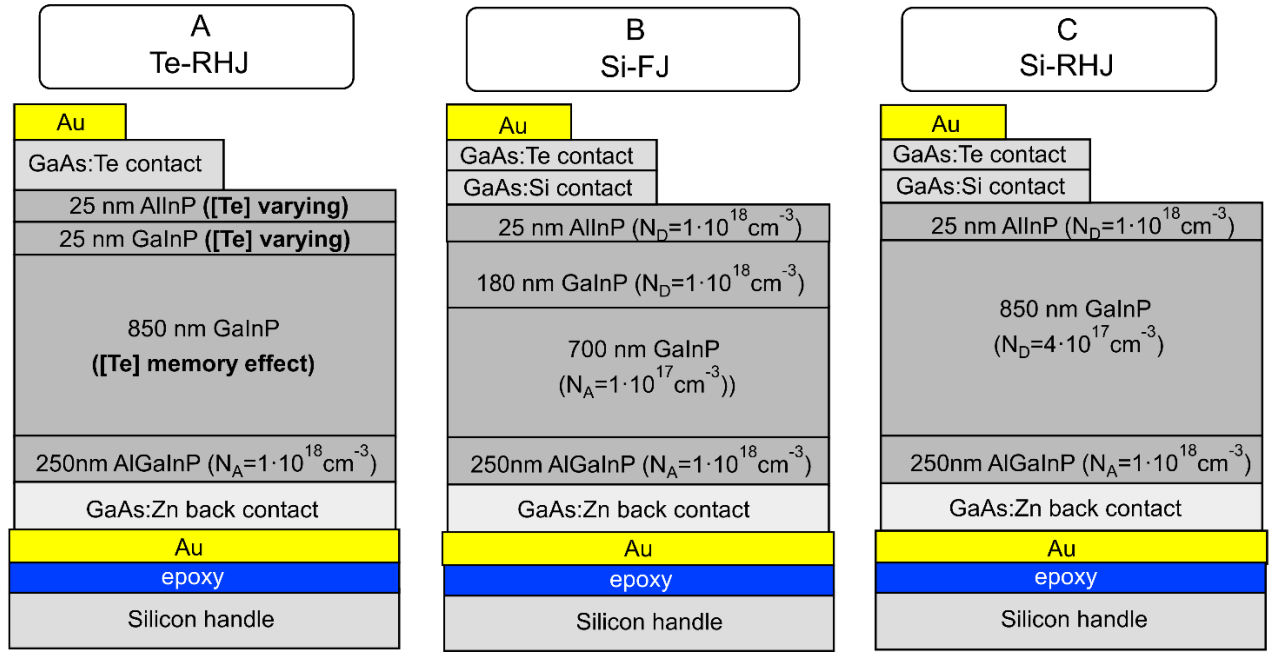

**Figure 2.** Structures of the GaInP solar cells compared in this work (a) Te-RHJ; (b) Si-FJ and (c) Si-RHJ. In (a), the n-type doping of a thick absorber is obtained through Te memory effect using four variants of Te preload in the contact layer and first nanometers of the phosphide layers.

These mentioned Te-RHJ structures are compared to two benchmark GaInP solar cells: a traditional front-junction design consisting of a thin-emitter of 180 nm Si-doped GaInP and a 700 nm Zn-doped GaInP base (Si-FJ); and a RHJ based on a 850 nm Si-doped thick GaInP emitter (Si-RHJ). Further details of the structures are provided in Fig. 2. All solar cells were fabricated using the inverted metamorphic (IMM) solar cell fabrication process as described in [19] with an active area of 0.1 cm$^2$ and an inverted square front grid. Both front and rear contacts are based on ~300 nm of gold deposited by electroplating. Concentrator solar cells are also fabricated in the same way, but using a Pd/Ge/Ti metal system for the front contact, leading to a specific semiconductor/metal resistance of $10^{-6}$ Ω·cm$^2$ and a metal sheet resistance around 30 mΩ/□ [18], minimizing the impact of the front grid on the performance at high concentrations. Electrical doping profiles of different samples were obtained by electrochemical capacitance voltage (ECV). Compositions were measured by conventional high-resolution X-Ray diffraction (XRD) scans, obtained with an X-Pert Panalytical diffractometer. Electroluminescence (EL) measurements were taken using a Keithley 2602A instrument for current bias and a Maya2000, fiber-based, calibrated spectrometer for light detection. Cross-sectional cathodoluminescence (CL) measurements were taken using a XiCLOne (Gatan UK) CL system attached to a LEO 1530 (Carl-Zeiss) field-emission scanning electron microscope (FESEM), using e-beams of 5 kV. The excitation volume in the sample is simulated by a Monte Carlo simulator (CASINO) [20]. Solar cell characterization includes external quantum efficiency (EQE) and reflectance (R) carried out using a custom-made system based on a Xe lamp and grating monochromator. The internal quantum efficiency (IQE) was calculated from the EQE and the R as IQE=EQE/(1-R). Dark and one-sun I-V curves were taken using a Keithley 2602A instrument and the light source was a Xe-lamp based solar simulator. The bandgap offset ($W_{oc}$) was calculated as $W_{oc}=E_g/q–V_{oc}$. $E_g$ is estimated as the wavelength corresponding to the maximum emission in the EL spectra. Emitter sheet resistances were obtained using the Van der Pauw method [21]. The solar cell devices used for EQE measurements have front contacts without grid to eliminate shadowing on the measurement. Finally, I-V curves under concentration were measured using a custom made, flash-lamp based, setup.

## 3. Results and Discussion

### 3.1. Material characterization

All growth conditions (except the injection of Te during the preload) are kept constant in order to evaluate the effect of Te on the electrical and optical properties of GaInP independently of other parameters. By using the same substrate orientation, V/III ratio, phosphine partial pressure and growth rate –all known to alter the degree of ordering in GaInP [22] [23] [24]–, the surfactant activity of Te can be correlated with an energy bandgap variation, for a given composition. However, although all samples were grown nominally lattice-matched to the GaAs substrate, deviations in the order of ±200 arcsec were observed in XRD scans, revealing slight deviations in composition, strain and, therefore, in the bandgap. In order to obtain fair comparisons, fully relaxed and perfectly lattice-matched to GaAs bandgaps are also calculated by using the method proposed in [25] and assuming fully strained GaInP layers. Therefore, the characteristic parameters contemplated for each sample in this section, presented in Table I, include: (1) phosphide Te preload flow; (2) energy bandgap of the resulting sample; (3) angular separation of the substrate and layer peaks in XRD; (4) composition and, finally, (5) corrected bandgap for an exactly lattice-matched composition.

**Table I.** Bandgap obtained by EL; layer and substrate XRD peaks angular separation; Ga composition; and corrected bandgap of different solar cells grown by using different preload DETe molar flows ($f_{DETe}$)

| Sample ID | $f_{DETe}$ preload flow (mol/min) | $E_g$ (eV) | XRD angular separation (arcsec) | Ga composition | Corrected $E_g$ (eV) |
|---|---|---|---|---|---|
| Te-RHJ#1 | - | 1.835 | 164 | 0.522 | 1,831 |
| Te-RHJ#2 | $5 \cdot 10^{-9}$ | 1.873 | -65 | 0.511 | 1.876 |
| Te-RHJ#3 | $4 \cdot 10^{-8}$ | 1.871 | -92 | 0.510 | 1.875 |
| Te-RHJ#4 | $3 \cdot 10^{-7}$ | 1.874 | -205 | 0.504 | 1.883 |

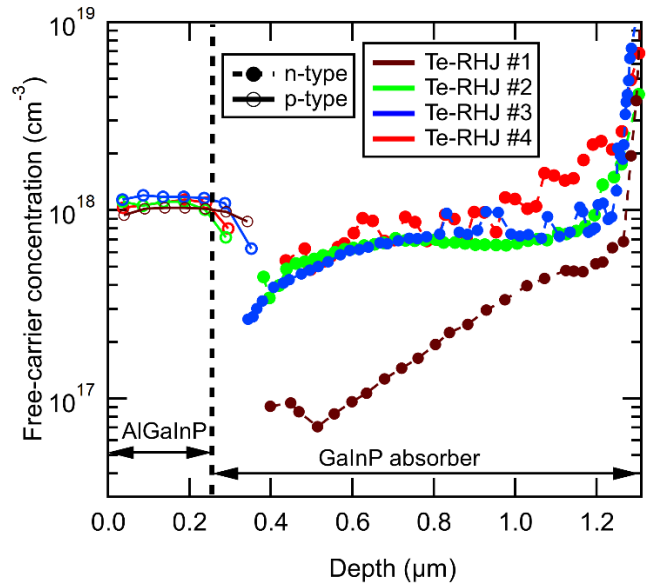

**Figure 3.** ECV measurements taken from the AlGaInP (p-type) layer to the GaInP absorber (n-type) of the 4 different solar cells grown using different DETe molar flows ($f_{DETe}$) at the initial 50 nm of the AlInP/GaInP layers.

The memory effect of Te gives rise to a gradual free-electron concentration along the GaInP layer in all cases (Fig.3) as expected from previous works [8]. However, the doping level along the total thickness, as well as the bandgap, get higher with higher phosphide Te preloads. Bandgap values as high as ~1.88 eV were measured. Surprisingly, Te-RHJ#2 and Te-RHJ#3, with one order of magnitude different phosphide Te preload flows, present similar carrier concentration profiles, whereas the doping profile is significantly altered when the DETe phosphine preload is increased an additional order of magnitude (Te-RHJ#4). This case results in a much higher electrical doping level at the beginning of the absorber which eventually converges to the same value of Te-RHJ#2 and Te-RHJ#3 by around the middle of the GaInP layer thickness. In Te-RHJ#1 there is no injection of Te during the phosphide layers, so the presence of Te in these layers is directly attributed to the memory effect of the Te injected during the growth of the GaAs contact layer. On the other hand, Te-RHJ#2, Te-RHJ#3 and Te-RHJ#4 sweep different Te preloads during the first nanometers of phosphides growth, apart from that already coming from de contact layer. The fact that the total amount of Te moles injected during the growth of the GaAs contact layer is always more than 10 times higher than the total Te injected in the AlInP/GaInP phase, but, at the same time, the doping profile changes abruptly when a phosphide Te preload flow is used (Te-RHJ#1, without Te phosphide preload, exhibits much lower average donor concentration than the others) suggests that the Te memory effect is much stronger in phosphide than arsenide layers. This is in accordance with the SIMS profile presented in Fig. 1, which shows a rapid decline in [Te] during the growth of the GaAs contact layer as soon as the Te injection is shut-off, but [Te] spikes back up and then slowly decreases during the phosphide layers growth.

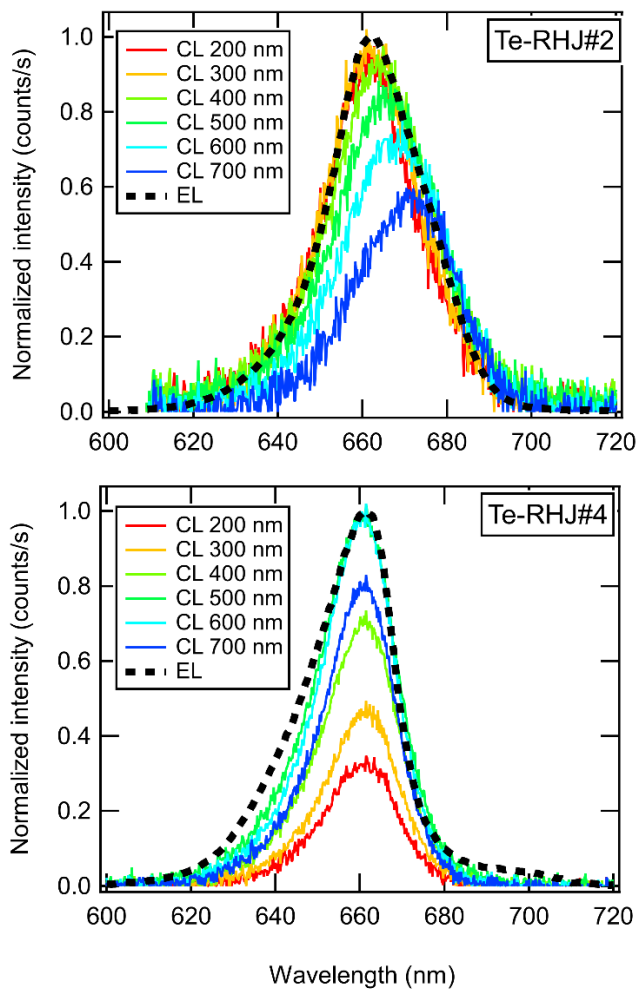

**Figure 4.** EL and cross-sectional CL along the GaInP absorber depth for samples Te-RHJ#2 (top) and Te-RHJ#4 (bottom). The depths in the legend for CL scans are measured from the AlInP/GaInP interface.

On another note, the Te memory effect is expected to give rise to gradual properties along the GaInP absorber layer depending on its initial injection, as it occurs in the doping profile. Since the concentration of Te on the growth surface presumably changes as it is removed from the growth chamber as well as it is being incorporated into the bulk, it might lead to a reduction on its surfactant activity, affecting the bandgap of the GaInP material grown. To further investigate a possible bandgap grading, normalized plan-view EL and cross-sectional spatially resolved CL emission spectra of Te-RHJ#2 and Te-RHJ#4 are compared in Fig. 4. The samples are chosen to represent the extreme cases of Te phosphide preload explored in this study. In EL and cross-sectional CL techniques, the samples are excited using two different approaches: in EL, current is injected through the electrical contacts of the fabricated solar cell devices, so the emitted light corresponds to the fraction of the total current injected into the device which recombines radiatively across the whole volume of the absorber layer. On the contrary, in CL the sample is excited by a 5 kV e-beam and the spectra are obtained by the emission in a locally excited volume, with a lateral full-with at half-maximum (FWHM) less than 20 nm according to simulations. Therefore, these locally emitted spectra allow to spatially resolve the material bandgap across the GaInP absorber layer as a function of depth.

A first inspection reveals that the maximum in the EL spectra is located around 660 nm (1.87 eV) in both samples, regardless of the initial Te injected. Differences are found on the shape of the EL spectra, though: Te-RHJ#4, with a two orders-of-magnitude higher Te preload presents a significantly narrower band emission than Te-RHJ#2. CL spectra along the GaInP layers provide insight on this point. The shape and position of the different peaks remains unaltered along the whole GaInP absorber layer in Te-RHJ#4, whereas a progressive redshift along the layer is produced from 1.87 eV to 1.84 eV in Te-RHJ#2. Note that, if variations as high as ~ 30 meV in layers thicker than 200 nm were induced by variations of the composition, high angular separations above 500 arcsec should be visible in the XRD scans. Therefore, the observed gradual bandgap may be explained by a progressive increase in the ordering degree as the concentration of tellurium on the growth surface fades away. In contrast, in Te-RHJ#4, the preload is high enough to cause a strong and lasting memory effect that does not fade away fast enough during the growth of the GaInP absorber layer; so the effect on the ordering of the material and on its bandgap remains constant.

Overall, surfactant properties of Te are proven through a dependence between the phosphide Te preload flows, the doping profile and a GaInP bandgap increase caused by reduction of the ordering degree. Taking into account the corrections for the slight deviations from the lattice-matched compositions, the bandgap shift goes from 1.831 up to 1.883 eV (~52 meV), pointing out Te as an electrically active impurity and a surfactant capable of partially disordering the material, presenting a heavy memory behavior in both cases.

### 3.2. Solar cells characterization

The IQE measured are shown in Fig. 5. Te-RHJ solar cells present lower cut-off wavelengths than the benchmark samples, which employ Si and Zn as doping elements (Fig.5 top), due to the bandgap increase. The IQE response in Te-doped samples is improved in the whole wavelength range as the phosphide Te preload is reduced from $3\cdot10^{-7}$ to $4\cdot10^{-8}$ mol/min (from Te-RHJ#4 to Te-RHJ#3), reaching collection efficiencies similar to those obtained in Si-RHJ, but does not improve further when the flow is lowered below this level: responses of Te-RHJ#3 and Te-RHJ#2 are virtually identical. All RHJ devices present slightly lower collection efficiencies as compared with Si-FJ, regardless of employing Te or Si in the thick emitter. The reason lies on a more demanding minority carrier diffusion length required to obtain the same IQE in RHJ designs than in FJ architectures [26], owing to the absorption profile generated in a GaInP layer, with most light absorbed within the first nanometers. Specifically, more than half of the total photons of wavelengths lower than 600 nm are absorbed in the first 180 nm corresponding to the emitter thickness of the Si-FJ. Since photogenerated carriers must reach the junction before being collected, the average distance to cover is significantly shorter in this configuration than in RHJs.

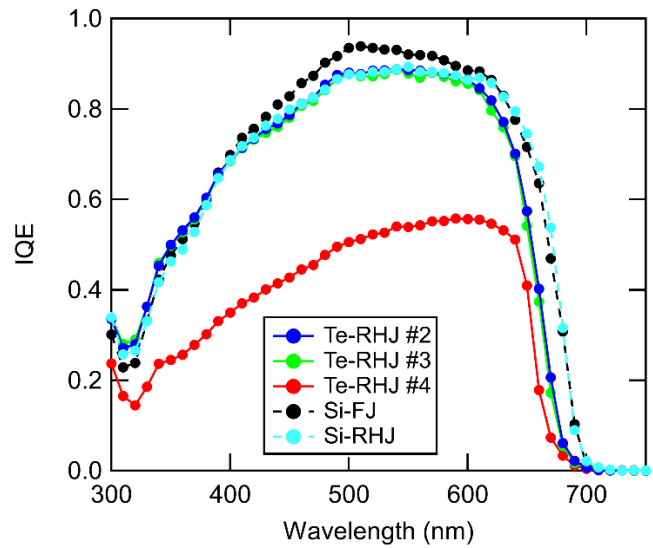

**Figure 5**. Measured internal quantum efficiencies (IQE) of tellurium-based rear-heterojunctions (Te-RHJ), traditional front-junction (Si-FJ) and silicon-based rear-heterojunctions (Si-RHJ).

In order to provide further insight on the dependence between the junction position and the collection efficiency, Fig.6 presents the modelled IQE of both configurations – FJ and RHJ – using the Hovel method [27] and the optical data used corresponding to our measured disordered-GaInP material [28]. To strictly focus on the junction position, the total thickness of GaInP is always 850 nm and the minority carrier properties –diffusion length ($L_m$) and lifetime ($\tau_m$)– are kept constant for both n and p-type GaInP. These parameters are extracted from the fit of $L_m$ to the measured IQE of Te-RHJ#2, assuming uniform properties in the absorber layer (which is not exactly the case in our gradual emitter cells but does not affect the purpose and conclusions of this modeling) with a fixed $\tau_p = 1$ ns [29][30] and a front surface recombination in the AlInP/GaInP interface $S_p = 100$ cm/s [31]. Besides, since almost all photons have already been absorbed in the emitter, the absorption in the AlGaInP base is negligible and it does not contribute significantly to the IQE in the modeling. Finally, the rear recombination velocity ($S_n$) in the base of the FJ, accounting for recombination in the GaInP/AlGaInP interface, is set to $10^5$ cm/s [31]. On the other hand, the GaInP/AlGaInP interface in the RHJ configuration is located inside the pn junction, where the action of the strong electric field sweeps the carriers across the interface, so this interface recombination is neglected. All parameters are summarized in Table II, where the suffix p refers to minority hole properties in the n-doped layer –emitter– and the suffix n refers to minority electron properties in the p-doped layer– base.

**Table II.** Parameters used in the Hovel modelling of both FJ and RHJ configuration.

|  | **RHJ** | **FJ** |
|---|---|---|
| Emitter/base | GaInP/AlGaInP | GaInP/GaInP |
| $x_{emitter}/x_{base}$ (nm) | 850/250 | 180/670 |
| $L_p/L_n$ (nm) | 1500/0.1 | 1500/1500 |
| $\tau_p/\tau_n$ (ns) | 1/1 | 1/1 |
| $N_D/N_A$ (cm$^{-3}$) | $5\cdot 10^{17}/1\cdot 10^{18}$ | $1\cdot 10^{18}/1\cdot 10^{17}$ |
| $S_p/S_n$ (cm/s) | Variable/$10^7$ | Variable/$10^5$ |

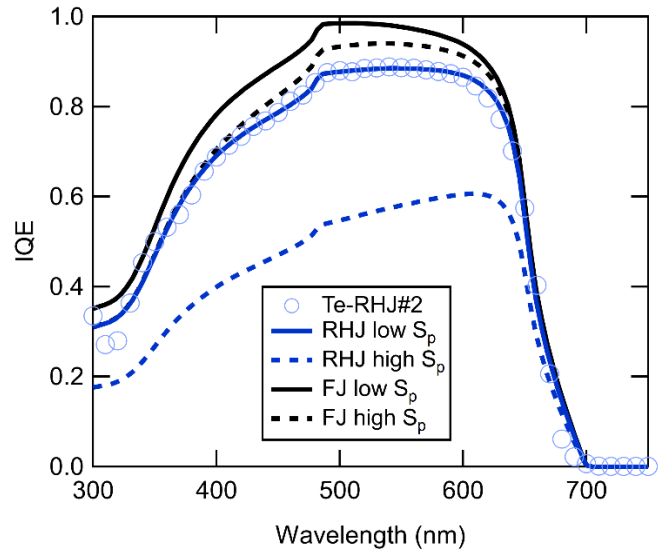

**Figure 6**. Modeled IQE of both front-junction and rear-heterojunction configurations. Recombination parameters are taken from the fit to the experimental response of Te-RHJ#2. Situations corresponding to a low (100 cm/s) and a high ($2.5\cdot 10^5$ cm/s) front surface recombination velocity are compared.

Therefore, the modeling results shown in Fig. 6 compare the experimental IQE corresponding to a RHJ design and the modeled IQE of a FJ design with the same parameters. In line with previous modeling work by Lumb et al. [26], Fig. 6 reveals a more favorable carrier collection in FJ than in RHJ designs, in the low $S_p$ case (100 cm/s), with peak values of 0.98 (instead of 0.88) and a 11% higher $J_{sc}$ (15.29 mA/cm$^2$ vs 13.73 mA/cm$^2$). Additionally, differences between collection efficiencies in both configurations become more pronounced at high $S_p$ (2.5·10$^5$ cm/s): in a thicker emitter the carriers have a higher probability of reaching and interacting with the front window-emitter interface before being collected at the junction. Therefore, a higher density of impurities, defects or lower local lifetimes due to high doping levels in the vicinity of this interface – accounted for by a high $S_p$ – could explain the deterioration of IQE in Te-RHJ#4, where a high flow of Te is directly injected during the growth of the window and the immediate subsequent 25 nm of GaInP.

Light and dark I-V curves measured on the solar cell devices under study are shown in Fig. 7. In Table III, the electrical parameters extracted from these curves, and Van der Pauw measurements for the emitter sheet resistance, are summarized. RHJ devices present lower $J_{sc}$, but at the same time, result in significantly better $V_{oc}$ and $W_{oc}$ in comparison with the traditional Si-FJ. Focusing first on the $J_{sc}$, two different contributions to the lower current can be differentiated in Te-RHJ samples, that can be understood with the discussing of the IQE presented above: the pn junction position, with a direct impact on the collection efficiency, and the bandgap increase, which reduces the total photons absorbed. On the one hand, differences of 0.35 mA/cm$^2$ in the $J_{sc}$ (14.98 mA/cm$^2$ vs 14.63 mA/cm$^2$) between Si-FJ and Si-RHJ, with very similar bandgaps, can be explained by the rear-junction position itself. On the other hand, the drop of 0.97 mA/cm$^2$ in the best Te-RHJ with respect to Si-RHJ (14.63 mA/cm$^2$ vs 13.66 mA/cm$^2$), with similar collection efficiencies but different bandgaps, can be attributed to bandgap variations. In this case, the impact of the bandgap (6.6%) is clearly more important than the change in the junction position (2.3 %). Nonetheless, the total thickness of Te-RHJ devices is still not optimized in the proof-of-concept solar cells presented in this work and a tradeoff between the carrier collection and the absorption arises from the emitter thickness in RHJs. While a thickness reduction leads to a lower photoabsorption, it may result beneficial for the overall $J_{sc}$, since it reduces the effective distance required to collect photogenerated carriers, enhancing the collection efficiency. In fact, preliminary modelling, not shown for brevity, reveals a $J_{sc}$ increase up to 14.07 mA/cm$^2$ in Te-RHJ#2 by simply thinning the emitter down to 650 nm.

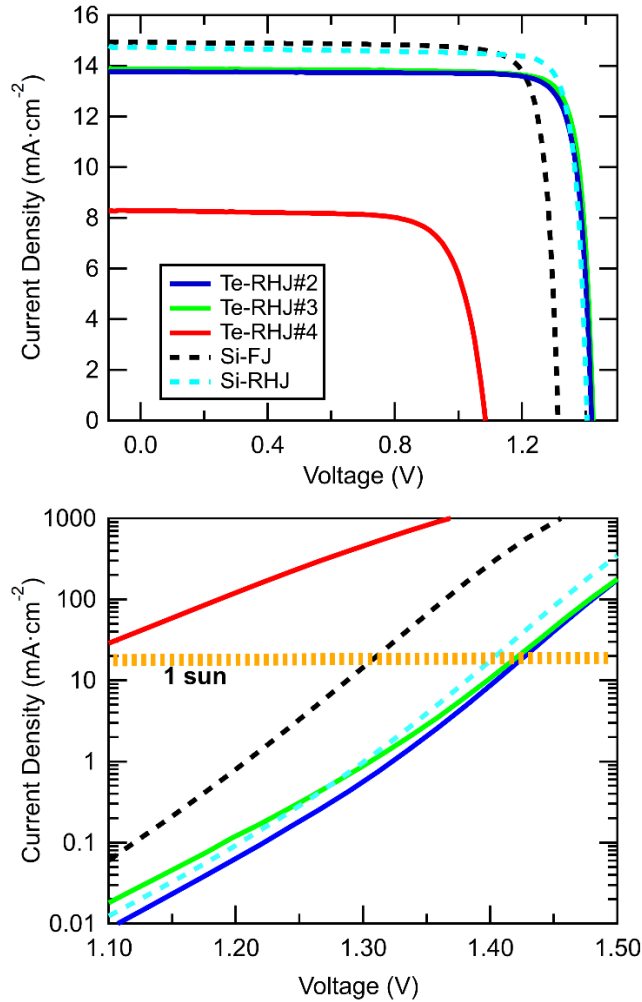

**Figure 7**. Light I-V (top) and dark I-V (bottom) curves of tellurium-based rear-heterojunctions (Te-RHJ), traditional front-junction thin-emitter (Si-FJ) and silicon-based rear-heterojunctions (Si-RHJ).

In a similar approach, two additive contributions to the $V_{oc}$ gain observed for Te-RHJ samples can be distinguished: the junction configuration and the bandgap shift. On the one hand, the $V_{oc}$ in Si-RHJ is 91 mV higher than in Si-FJ (~6.9%), with similar bandgaps. On the other hand, among all the RHJ samples, Te-RHJ#3 exhibits the highest $V_{oc}$, 1.422 V (followed very closely by Te-RHJ#2 with 1.418 mV), representing an additional 18 mV to the Si-RHJ voltage (~1.3%). This difference is exclusively attributed to the higher bandgap achieved by a higher disorder. Naturally, the aggregate $V_{oc}$ increase becomes more pronounced when comparing the best Te-RHJ to the traditional

Si-FJ (1.422 vs 1.313V). In contrast to the case of the $J_{sc}$, the junction configuration is more influential than the bandgap increase on the recombination current and $V_{oc}$. In fact, the lower $W_{oc}$ in Si-RHJ than in Te-RHJ reflects this idea.

Table III. Electrical parameters of the different GaInP solar cells studied. The short-circuit current has been calculated from the IQE curve using the AM1.5D G173 direct spectra.

| Structure | $f_{DETe}$ preload (mol/min) | $J_{sc}$ (mA/cm$^2$) | $V_{oc}$ (V) | $E_g$ (eV) | $W_{oc}$ (V) | $R_{she}$ (Ω/□) | FF |
|---|---|---|---|---|---|---|---|
| **Te-RHJ#2** | 5·10$^{-9}$ | 13.66 | 1.418 | 1.873 | 0.460 | 147 | 0.87 |
| **Te-RHJ#3** | 4·10$^{-8}$ | 13.47 | 1.422 | 1.871 | 0.464 | 152 | 0.87 |
| **Te-RHJ#4** | 3·10$^{-7}$ | 8.63 | 1.085 | 1.874 | 0.858 | 127 | 0.76 |
| **Si-FJ** | - | 14.98 | 1.313 | 1.838 | 0.539 | 551 | 0.85 |
| **Si-RHJ** | - | 14.63 | 1.404 | 1.841 | 0.448 | 320 | 0.87 |

Since many cross-related parameters are modified at the same time –doping levels, type of minority carriers in the absorber, bandgap or the use of AlGaInP in the junction– the use of the external luminescence efficiency, $\eta_{ext}$, as a figure of merit to compare the different solar cells is highly convenient [32][33][34]. This parameter accounts for the portion of the total recombination current (equal to the injected current in the dark, $J_{inj}$) which is radiative ($J_{0rad}/J_{inj}$) and determines how close the device is to the radiative limit [35]. In Fig.8, $\eta_{ext}$ is represented, extracted from EL measurements at different $J_{inj}$ of Te-RHJ#2, Si-RHJ and Si-FJ samples.

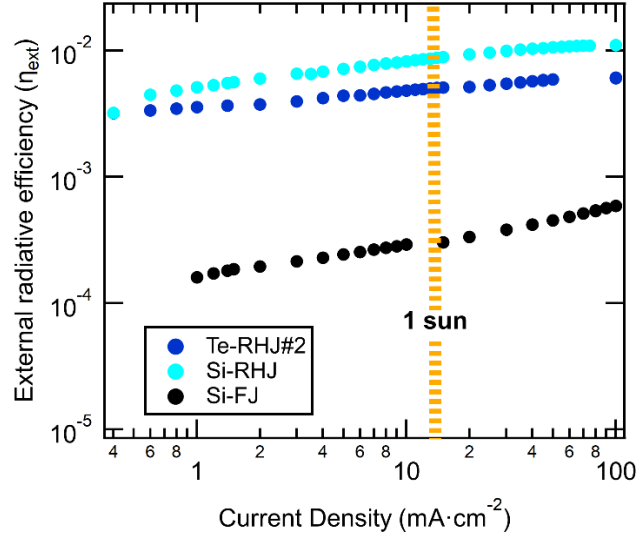

**Figure 8**. Measured external radiative efficiency of three representative samples: tellurium-based rear-heterojunction (Te-RHJ#2), traditional front-junction (Si-FJ) and silicon-based rear-heterojunction (Si-RHJ).

A similar tendency as in [14] is observed in our samples: all RHJs have an order-of-magnitude higher radiative efficiencies than the traditional Si-FJ at current density values equivalent to ~1 sun, which explains the lower recombination currents shown in the dark I-V curves, and the lower $W_{oc}$. Such improvement may be originated by different factors. In [14] and [36] the authors coincide in pointing out a reduction of non-radiative Sah-Noyce-Shockley (SNS) recombination ($J_{0m}$) in the depletion region [37], as the main reason of the recombination decrease and $V_{oc}$ increase in a RHJ device. Reduction of SNS recombination is achieved by thinning the depletion region through the introduction of a high bandgap material in the junction (AlGaInP). However, $J_{0m}$ should stop dominating at high current densities because of the higher ideality factor ($n \sim 2$) than the bulk recombination current ($J_{01}$), with ($n \sim 1$). Therefore, the higher $\eta_{ext}$ of RHJ at high current densities suggests that other factors may be contributing to lowering non-radiative recombination, an issue which is still under investigation. In Fig. 8, it can also be observed that the Si-RHJ sample has a higher $\eta_{ext}$ than the Te-RHJ, which is in accordance with the $W_{oc}$ obtained and indicates that the overall electronic quality of the Te-RHJ cells under development has room for improvement. Anyway, this lower $\eta_{ext}$ is overcome by a higher bandgap, leading to a higher $V_{oc}$ in Te-RHJ devices, as shown above.

### 3.3. Concentration performance

Achieving a very low front metal contact resistance, enabled by the use of Te as dopant, is of particular importance for concentrator applications. Additionally, in the Te-RHJ solar cells developed, a considerable decrease in the emitter sheet resistance was also obtained, from 551 $\Omega/\square$ in Si-FJ to 147 $\Omega/\square$ in Te-RHJ#2. This has important implications in the design and performance of the solar cell. On the one hand, it can be used to reduce the front grid shadowing factor and achieve a higher $J_{sc}$. On the other hand, it can be used to improve the performance at higher concentrations. To assess the impact of these improvements, combined with the lower $J_{sc}$ and higher $V_{oc}$ obtained, on the performance of the GaInP cell developed, we have carried out concentration response measurements on the traditional cell with thin emitter (Si-FJ) and the gradual RHJ cell presented in this work (Te-RHJ#2). To aid in the analysis, we have modelled the concentration response as well. For this we have used our proven model based on distributed circuit units [38] [39] [40] with the measured short circuit currents, recombination currents, ideality factors and emitter sheet resistances as experimental input parameters (see Table III and Fig. 7). The solar cells simulated reproduce the size and geometry of the experimental cells developed: 0.1cm$^2$ active area with an inverted square front grid geometry and 4 µm wide fingers, similar to our typical concentrator solar cell design [41][42]. The metal and contact parameters used correspond to our state-of-the-art metallization for inverted solar cells, which relies on a high doping level in the contact layer achieved using tellurium [18], and were confirmed using TLM and Hall van der Pauw measurements. The results of the simulations and measurements are shown in Fig. 9. Note that the concentration in the X-axis refers to times the 1-sun $J_{sc}$ of each cell measured.

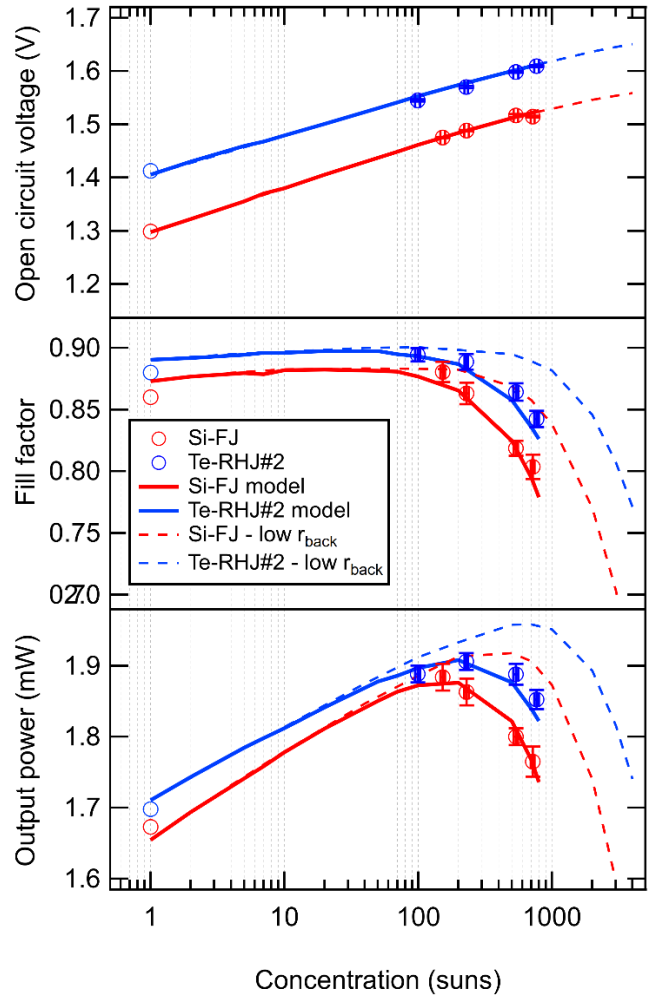

**Figure 9**. Symbols: concentration measurement results for Si-FJ and Te-RHJ#2 GaInP cells studied. The values shown are the average of a set of measurement, and the error bars represent the standard deviation in these (note that horizontal error bars are present too but difficult to distinguish). Solid lines: fit to these measured concentration responses. Dashed lines: modeled concentration response obtained using the same fitting parameters and a lower back contact resistance in the inverted solar cells. Distributed modeling based on electronic circuit units is used with the experimental parameters obtained (Table III and Fig. 7) and the front grid parameters of our state-of-the-art metallization process, as input data.

The concentration response of the $V_{oc}$ shows the expected logarithmic behavior, with a slightly different slope for the two cells due to the different ideality factors observed also in the dark I-V curves (Fig. 7). The model agrees very well with the experimental data points, as can be seen, and shows that the $V_{oc}$ gain obtained with the gradual emitter design decreases only slightly with concentration, consistently with the small difference in ideality factors. The FF results are clearly better under concentration for the case of the RHJ design, as expected. In these cells, the FF peaks at around 200 suns and then starts a fast decrease, despite the front grid is optimized for higher concentrations. This is caused by a high series resistance originated at the back of the prototype inverted solar cells implemented, which have a thin back contact metal layer and are glued to a silicon handle by means of a non-conductive epoxy (see Fig 2). The dashed lines in Fig. 9 correspond to simulations carried out using the same model parameters but a design of the cell where there is no epoxy at the back. This case shows a better FF response under concentration, peaking at higher concentrations. Finally, the output power measurement allows to assess the combined effect of $J_{sc}$, $V_{oc}$ and FF. Note that the active area of the solar cells measured are virtually identical, so the direct comparison of output power is appropriate. It can be seen that a significantly higher power is obtained with the RHJ design. The advantage is higher at higher concentrations.

Therefore, we can conclude that the lower $J_{sc}$ obtained in the Te-based RHJ design is compensated for by the higher $V_{oc}$ and FF, achieving higher efficiencies at 1-sun and under concentration. The results presented here are intended to be understood as proof of concept, with clear room for improvement, particularly in the case of the $J_{sc}$ of the solar cells. The applicability of these GaInP top cells to enhance the efficiency of multijunction solar cells will require improving the carrier collection efficiency and $J_{sc}$, at least for series connected subcell devices. As standalone solar cells, they could already exhibit better performances than the traditional design in applications such as power harvesters for the Internet of Things (IoT) [43], multi-terminal mechanically stacked cells [44], or recently developed 3-terminal heterojunction bipolar transistor cells [45]. The growth simplicity brought about by the use of just one n-type dopant in the Te-RHJ design can be an interesting advantage too. Nevertheless, realizing the potential improvements in the carrier collection efficiency discussed will also be of benefit to all these applications.

## 4. Conclusions

We have demonstrated a method to grow an inverted rear-heterojunction GaInP solar cell with an absorber doping profile which relies exclusively on the memory effect of Te coming from the heavily doped front contact layer. This way, a low front metal contact resistance can be achieved while using to our advantage the otherwise detrimental memory effect of Te. This strategy requires the understanding of the incorporation of Te in the different III-V layers during the epitaxial growth and takes advantage of its properties as surfactant, which allows to reduce the GaInP CuPt ordering thus increasing its bandgap. This fact together with the use of a RHJ architecture results in an important $V_{oc}$ and $W_{oc}$ enhancement in comparison with the traditional thin-emitter Si-FJ design. The results have been analyzed through the comparison of the QE and I-V curves of a Te-RHJ, Si-RHJ and a thin-emitter Si-FJ. In addition, this approach also gives rise to a notable reduction in the emitter sheet resistance, which enables reducing the front grid shadowing factor for an increased $J_{sc}$ and/or a better concentration response. The RHJ solar cells developed so far exhibit a slight decrease in the collection efficiency and $J_{sc}$, which can be improved by a redesign of the emitter thickness. We have shown that, overall, the RHJ solar cells developed can achieve an improved conversion efficiency and enhanced concentration response when processed as concentrator solar cells. Ongoing work pursues further improvements by redesigning the gradual doping profile.

## Acknowledgments

This project has been funded by the Spanish MINECO with the project TEC2017-83447-P, by the Comunidad de Madrid with the project with reference Y2018/EMT-4892 (TEFLÓN-CM) and by Universidad Politécnica de Madrid by Programa Propio. M. Hinojosa is funded by the Spanish MECD through a FPU grant (FPU-15/03436) and I. García is funded by the Spanish Programa Estatal de Promoción del Talento y su Empleabilidad through a Ramón y Cajal grant (RYC-2014-15621). The authors want to thank Norman R. Jost for helping with the electroluminescence measurements and Shabnam Dadgostar and the Optronlab group for their technical support with the cathodoluminescence measurements.